\documentclass{article}

\usepackage{amsfonts}
\usepackage{amsmath}
\usepackage{graphicx}
\begin{document}

\begin{center}
\Large
THE QUANTUM GEOMETRIC TENSOR FROM GENERATING FUNCTIONS\\ 
J. Alvarez-Jim\'enez \footnote{E-mail: {\tt javier.alvarez@correo.nucleares.unam.mx}} and J. D. Vergara\footnote{E-mail: {\tt vergara@nucleares.unam.mx}}\\
\normalsize
{Instituto de Ciencias Nucleares, \\
Universidad Nacional Aut\'onoma de M\'exico, A. Postal 70-543, \\
M\'exico CDMX, M\'exico.}
\end{center}

%\date{}
%\maketitle

\begin{abstract}
We introduce a new method to compute the Quantum Geometric Tensor, this procedure allow us to compute the Quantum Information Metric and the Berry curvature  perturbatively 
for a theory with an arbitrary interaction Hamiltonian. The calculation uses the generating function method, and it is illustrated with the harmonic oscillator with a linear and a quartic perturbation.
\end{abstract}

\section{Introduction}	 

For some years, several notions of statistical distance have been worked on, among the first correspond to Fisher's distance \cite{Fisher},  this notion of statistical distance has had countless applications, between them its introduction in the quantum mechanical context, see for example \cite{Wootters}.  To give a geometric meaning to this distance, the idea of a gauge invariant metric on a projective Hilbert space was later introduced by Provost and Vallee in  \cite{Provost}. This metric in a mathematical context corresponds to the Fubini-Study metric \cite{Kobayashi, Abe}, and it has been extensively studied in the field of quantum information theory, see for instance \cite{Amari, Fujiwara}. Furthermore,
in recent years, a lot of work has been done in order to understand the role that information theory plays in Quantum Mechanics, and it has been found that within this theory there is a rich geometric structure\cite{Provost,EGQS,GSprobst,phest,zanardi}, where play role several structures. One of these approaches consists in defining the Quantum Geometric Tensor (QGT) of the parameter space of the system\cite{Provost}.  This tensor contains as a real part the Quantum Information Metric (QIM), that is equivalent to the Fubini-Study metric, and the imaginary part corresponds to the Berry's curvature, whose flux gives the Berry phase\cite{Berry}. Geometrical phases have several applications for example topological quantum computation \cite{review}, they also control a crucial effect in quantum mechanics,  quantum interference \cite{13}, and have also been studied for mixed states\cite{geophasmix}. A different, but  related tool of quantum information, is the Quantum Fidelity. The Quantum Fidelity is an essential tool to predict quantum phase transitions\cite{fifidphtr,Gu} and the Fidelity susceptibility \cite{Gu}, corresponds essentially to the QIM. Some approaches to these geometrical structures involve the path integral for computing the Quantum Fidelity\cite{pathfidel},  the QIM  \cite{MIyaji:2015mia,Trivella:2016brw} and the QGT\cite{paper}. 
%A interest in field theory\cite{Higgs}

 In this paper, we show that the QGT can be computed by using quantum field theory techniques. The standard approach when  a problem  cannot be solved exactly is  to apply perturbation theory  directly to compute the wavefunction. However, this can sometimes be very hard or even impossible (take for instance quantum field theory). Fortunately, in quantum field theory the perturbation theory is widely applied to compute Green's functions of the theory and we want to develop this technique to obtain a perturbative form of calculating the  Quantum Geometric Tensor. When the problem is not exactly solvable, the procedure allows us to calculate the QMT perturbatively even when the wave function is completely unknown. This procedure is illustrated by computing the  QGT for a harmonic oscillator, with a linear perturbation in an exact way. Furthermore, we consider an anharmonic oscillator with a $q^4$ perturbation. In this case, the QMT is computed to first order in the coupling constant.

The structure of this paper is as follows. In Section 2, we introduce our procedure to built the Quantum Geometric Tensor, following the approach developed in Ref. \cite{paper}, but specialized to the case of Quantum Mechanics. In this form, the construction is more transparent and reveals that it works in the case of the ground state of the full theory. In the sense that if we can compute exactly the Green's functions, our procedure will give us the exact result of the Quantum Geometric Tensor. Of course, there is a reduced number of examples of this case, but in Sec. 5.1 we show how the procedure works using an exactly computed Green's function. Section 3 indicates that the QGT can be computed perturbatively for an arbitrary potential $V(q)$,  to any order in the coupling constant, taking in mind that the value of the coupling be small.  Section 4, presents a concise description of the perturbative  approach that will be of interest for our proposes. In Section 5 we analyze two examples to clarify our method, and finally, in Section 6, we present our conclusions.

\section{The Quantum Geometric Tensor}

Let us assume  the system we are going to describe depends on $N$ parameters $\lambda^a$, where $a=1,...,N$. If we consider a small variation in the parameter space $\lambda \rightarrow  \lambda + \delta \lambda$ we can compute the overlap, $f(\lambda,\lambda+\delta\lambda)$, between the corresponding states $\left| \Psi(\lambda)\right>$ and $\left| \Psi(\lambda + \delta \lambda)\right>$
\begin{equation}
f(\lambda,\lambda+\delta \lambda)=\left< \Psi(\lambda) \right. \left|\Psi(\lambda+\delta \lambda)\right>.
\end{equation}

The Quantum Fidelity $F(\lambda,\lambda')$ is defined like the modulus of the overlap
\begin{equation}
F(\lambda,\lambda')= \left| f(\lambda,\lambda') \right|,
\end{equation}
we will consider the case $\lambda'=\lambda+\delta \lambda$, so we can expand near $\lambda$
\begin{equation} \label{eq:expfide}
F(\lambda,\lambda+\delta\lambda)= 1 - \frac{1}{2} G_{\alpha \beta} \delta\lambda^\alpha \delta\lambda^\beta + \mathcal{O}(\lambda^3), 
\end{equation}
where $G_{\alpha \beta}$ is the Quantum Geometric Tensor (QGT) \footnote{Some authors prefer to define the QGT absorbing the $1/2$ that appears in expansion (\ref{eq:expfide}). }
\begin{equation}
G_{\alpha \beta}= \left< \partial_\alpha \Psi \right. \left| \partial_\beta \Psi \right> - \left< \partial_\alpha \Psi \right. \left| \Psi \right>\left< \Psi \right. \left| \partial_\beta \Psi
 \right>.
\end{equation}
 In the expansion (\ref{eq:expfide}), it is clear that only the symmetric part of the Quantum Geometric Tensor will contribute to the fidelity. From its definition, we can see that the symmetric part coincides with the real part, this is known as the Quantum Information Metric (QIM) and is given by
 \begin{equation}
 g_{\alpha \beta}= \mathbf{Re} G_{\alpha \beta} = \frac{1}{2}\left( \left< \partial_\alpha \Psi \right. \left| \partial_\beta \Psi \right> +  \left< \partial_\beta \Psi \right. \left| \partial_\alpha \Psi \right> \right) - \left< \partial_\alpha \Psi \right. \left| \Psi \right>\left< \Psi \right. \left| \partial_\beta \Psi
 \right>.
 \end{equation}
 
 Even though the antisymmetric and imaginary part does not contribute to equation (\ref{eq:expfide}), it also has a physical meaning, it is proportional to the Berry curvature $F_{\alpha \beta}$
 \begin{equation}
 \frac{1}{2}F_{\alpha \beta}= \mathbf{Im} G_{\alpha \beta} = \frac{1}{2 i}\left(  \left< \partial_\alpha \Psi \right. \left| \partial_\beta \Psi \right> -  \left< \partial_\beta \Psi \right. \left| \partial_\alpha \Psi \right> \right).
 \end{equation}
 The above definitions are well known and can be used for any quantum state, but the critical point is that we need to know the explicit form of this state. However,  in the case of the ground state it is possible to follow an alternative route, see for example \cite{MIyaji:2015mia, Trivella:2016brw, paper}. In this case, we do not need the explicit form of the quantum state, and our starting point is the Lagrangian of the system. In the following section, we will show how this procedure works.
 
\subsection{The Quantum Geometric Tensor From a Lagrangian Approach}
 Let us suppose that  the system is described by the Lagrangian $L_i$ during the time interval $(-\infty, 0)$. Now, during the interval $(0,\infty)$ we add to the system a perturbation given by the modification of the parameters $\lambda_i \to \lambda_i + \delta\lambda_i$. Then the new Lagrangian will be $L_f=L_i +\mathcal{O}_i \delta \lambda^i$. With this in mind, we can write the amplitude $\left< q_f,t_f| q_i, t_i \right>$ by inserting an identity in the basis of the energy and we will get 
\begin{eqnarray}
\left< q_f,t_f | q_i, t_i \right> = \sum_{n} \left< q_f, t_f | n\right> \left< n | q_i, t_i \right> \nonumber \\ = \sum_{n,n'} \left< q_f\right| e^{- i t_f E_{n'}^f} \left| n' \right>\langle n' |n\rangle  \left< n \right| e^{it_i E_{n}^i} \left| q_i \right>,
\end{eqnarray}
where the subindexes 
  $i$ and $f$ refers to the Lagrangian that describes the theory. 
  Now we make a Wick rotation
\begin{equation}
t \longrightarrow -i \tau,
\end{equation}
with this change, the exponentials turn to
\begin{equation}
e^{-i t_f E_n^f} \longrightarrow  e^{- \tau_f E_n^f}, \ \ \   e^{i t_i E_n^i} \longrightarrow  e^{ \tau_i E_n^i},
\end{equation}
since we can always adjust $E_n^i$ and $E_n^f$ to be higher than zero, all the exponentials will tend to zero when we take the limits
\begin{equation}
\tau_f \longrightarrow \infty,  \ \ \ \tau_i \longrightarrow - \infty,
\end{equation}
and the ground state energies will dominate the summation, so we can keep only the ground state term
\begin{eqnarray}
\left< q_f,\infty | q_i,-\infty \right> &=& e^{- \tau_f E_0^f} e^{\tau_iE_0^i}  \left< q_f | 0_f\right> \left< 0_f | 0_i \right> \left< 0_i | q_i \right> \nonumber \\ &=& \left< q_f,\infty | 0_f\right> \left< 0_f | 0_i \right> \left< 0_i | q_i, -\infty \right> ,\end{eqnarray}
then we have
\begin{equation} \label{eq:traspsants}
\left< 0_f | 0_i \right> = \frac{\left< q_f,\infty | q_i,-\infty \right>}{\left< q_f,\infty | 0_f \right> \left< 0_i | q_i,-\infty \right>}.
\end{equation}
Let us now analyze each term. In the first we insert an identity in the form $\mathbb{I}=\int dq_0 \left| q^0 \right> \left< q^0\right|$

\begin{eqnarray} \label{eq:trasidcamc}
&&\left<  q_f, \infty| q_i,-\infty \right> = \int dq^0 \left< q_f,\infty | q^0 \right> \left< q^0 | q_i,-\infty \right>\nonumber \\
&=& \int dq^0 \int_{q(t=0)=q^0}^{q(\infty)} \mathcal{D}q e^{- \int_0^\infty d\tau' \mathcal{L}_f} \int_{q_i(-\infty)}^{q(t=0)=q^0} \mathcal{D} q e^{-  \int_{-\infty}^0 d\tau' \mathcal{L}_i}, \nonumber \\ &=& \int \mathcal{D}q \exp\left(-\int_{-\infty}^{\infty}d\tau  \mathcal{L}_i - \int_0^\infty d\tau  \delta \lambda^\alpha \mathcal{O}_\alpha\right).
\end{eqnarray}
\normalsize
We must clarify that in the above and the following expressions the Lagragian $\mathcal{L}$ corresponds to the Wick rotated Lagrangian.
For the denominator, we notice that
\begin{equation}
Z_f = \int \mathcal{D} q \exp\left(- \int_{-\infty}^\infty d\tau  \mathcal{L}_f \right),
\end{equation}
since the theory is time reversal invariant, we interpret
\begin{equation}
\left< q_f,\infty | 0_f \right> = \sqrt{Z_f},
\end{equation} from a similar argument we get
\begin{equation}
\left< 0_i | q_i,-\infty \right> = \sqrt{Z_i},
\end{equation}
therefore, the overlap takes the form
\begin{eqnarray} \label{eq:trasinttra}
\left< 0_f | 0_i \right> =   \dfrac{\int \mathcal{D} q \left( e^{ -\int_{-\infty}^\infty d\tau  \mathcal{L}_i } e^{-\int_{0}^\infty d\tau   \delta\lambda^\alpha \mathcal{O}_\alpha }\right)}{\sqrt{Z_i Z_f}}.
\end{eqnarray} 
In order to simplify, we use the definition of expectation value for an operator $A$:
\begin{equation}
\left< A \right> = \frac{1}{Z_i} \int \mathcal{D} q \left[ \exp \left( - \int_{-\infty}^\infty d\tau  \mathcal{L}_i\right) A(q) \right].
\end{equation}
Notice that we are defining the expectation value with respect to the initial Lagrangian $\mathcal{L}_i$.  In consequence the overlap of the vacuum states takes the form
\begin{equation}
\left< 0_f | 0_i \right>=\frac{\left< e^{-\int_0^\infty d\tau  \delta \lambda^\alpha \mathcal{O}_\alpha}\right>}{\sqrt{\frac{Z_f}{Z_i}}} = \frac{\left< e^{-\int_0^\infty d\tau  \delta \lambda^\alpha \mathcal{O}_\alpha}\right>}{\sqrt{\frac{\int \mathcal{D} q e^{-  \int_{-\infty}^\infty d\tau (\mathcal{L}_i + \delta \lambda^\alpha \mathcal{O}_\alpha)}}{Z_i}}},
\end{equation}
 from which we find that
\begin{equation} \label{eq:trasindtra}
\left< 0_f | 0_i \right> = \dfrac{\left< \exp\left( - \int_0^\infty d\tau  \delta\lambda^\alpha \mathcal{O}_\alpha(\tau)\right) \right>}{\sqrt{\left< \exp \left(- \int_{-\infty}^\infty d \tau  \delta \lambda^\alpha \mathcal{O}_\alpha(\tau) \right) \right>}},
\end{equation}
therefore
\begin{equation}
\left| \left< 0_f\right| \left. 0_i \right>\right|^2= \dfrac{\left< e^{\left(- \int_{0}^\infty d \tau  \delta \lambda^\alpha \mathcal{O}_\alpha(\tau) \right)} \right> \left<  e^{\left( -\int_{- \infty}^0 d \tau  \delta \lambda^\beta \mathcal{O}_\beta(\tau) \right)} \right>}{\left< \exp \left(- \int_{- \infty}^\infty d \tau  \delta \lambda^\rho \mathcal{O}_\rho(\tau) \right) \right>}.
\end{equation}
Expanding in series
\begin{align}
&\left| \left< 0_f | 0_i \right>\right|^2 = 1 + \left[  \dfrac{1}{2}\int_{0}^\infty d\tau_1 \int_{0}^\infty d \tau_2  \left< \mathcal{O}_\alpha(\tau_1) \mathcal{O}_\beta (\tau_2) \right> +\dfrac{1}{2}\int_{-\infty}^0 d\tau_1 \int_{-\infty}^0 d \tau_2  \left< \mathcal{O}_\alpha(\tau_1) \mathcal{O}_\beta (\tau_2) \right>  \right. \nonumber \\& - \dfrac{1}{2}\int_{-\infty}^\infty d\tau_1 \int_{-\infty}^\infty d \tau_2  \left< \mathcal{O}_\alpha(\tau_1) \mathcal{O}_\beta (\tau_2) \right>   \left. + \int_{0}^\infty d\tau_1 \int_{-\infty}^0 d \tau_2  \left< \mathcal{O}_\alpha(\tau_1) \right> \left< \mathcal{O}_\beta (\tau_2) \right> \right] \delta \lambda^\alpha \delta \lambda^\beta .
\end{align}
It should be noticed that the linear term vanishes. The next step to simplify  is only to notice that
\begin{eqnarray}
\int_{-\infty}^\infty d\tau_1 \int_{-\infty}^\infty d \tau_2 &=& \int_{-\infty}^0 d\tau_1 \int_{-\infty}^0 d \tau_2 +\int_{-\infty}^0 d\tau_1 \int_{0}^\infty d \tau_2 +\int_{0}^\infty d\tau_1 \int_{-\infty}^0 d \tau_2 \nonumber \\ &+& \int_{0}^\infty d\tau_1 \int_{0}^\infty d \tau_2,
\end{eqnarray}
since we  ask for time reversal symmetry
\begin{equation}
 \left< \mathcal{O}_\alpha(\tau_1) \mathcal{O}_\beta (\tau_2) \right> =  \left< \mathcal{O}_\alpha(-\tau_1) \mathcal{O}_\beta (-\tau_2) \right>,
\end{equation}
we find
\begin{equation}
\left| \left< 0_f | 0_i \right>\right|^2 = 1 -  \int_{-\infty}^0 d\tau_1 \int_{0}^\infty d \tau_2  \left[ \left< \mathcal{O}_\alpha(\tau_1) \mathcal{O}_\beta (\tau_2) \right>  \right.  \left. - \left< \mathcal{O}_\alpha(\tau_1) \right> \left< \mathcal{O}_\beta (\tau_2) \right> \right]\delta \lambda^\alpha \delta \lambda^\beta.
\end{equation}
Using the definition of the Quantum Fidelity, and taking into account that
\begin{equation}
F(\lambda,\lambda+\delta\lambda) = 1  -  \frac{1}{2}G_{\alpha \beta}\delta \lambda^\alpha  \delta \lambda^\beta,
\end{equation}
we obtain
\begin{equation} \label{eq:defQGT}
G_{\alpha \beta}  =   \int_{-\infty}^0 d\tau_1 \int_0^{\infty} d \tau_2 \left[  \left< \mathcal{O}_\alpha(\tau_1) \mathcal{O}_\beta(\tau_2)\right> \right.  \left. - \left< \mathcal{O}_\alpha(\tau_1) \right> \left< \mathcal{O}_\beta(\tau_2)\right> \right].
\end{equation}
 
The quantum information metric, which corresponds to the real part is
\begin{eqnarray} \label{eq:tmqintdetra}
 \textbf{Re} G_{\alpha \beta}  =   \textbf{Re} \int_{-\infty}^0 d\tau_1 \int_0^{\infty} d \tau_2 \left[  \left< \mathcal{O}_\alpha(\tau_1) \mathcal{O}_\beta(\tau_2)\right> \right.  \left. - \left< \mathcal{O}_\alpha(\tau_1) \right> \left< \mathcal{O}_\beta(\tau_2)\right> \right]  \\ =  \int_{-\infty}^0 d\tau_1 \int_0^{\infty} d \tau_2 \left[ \frac{1}{2} \left( \left< \mathcal{O}_\alpha(\tau_1) \mathcal{O}_\beta(\tau_2)\right> + \left< \mathcal{O}_\alpha(\tau_1) \mathcal{O}_\beta(\tau_2)\right> \right) \right. \nonumber \\  \left. - \left< \mathcal{O}_\alpha(\tau_1) \right> \left< \mathcal{O}_\beta(\tau_2)\right> \right] \nonumber \\  = \int_{-\infty}^0 d\tau_1 \int_0^{\infty} d \tau_2 \left[ \frac{1}{2} \left< \left\lbrace \mathcal{O}_\alpha(\tau_1), \mathcal{O}_\beta(\tau_2) \right\rbrace_+ \right>  - \left< \mathcal{O}_\alpha(\tau_1) \right> \left< \mathcal{O}_\beta(\tau_2)\right> \right] \nonumber ,
\end{eqnarray} 
whereas the Berry curvature  takes the form
\begin{eqnarray}
\textbf{Im} G_{\alpha \beta} &=& \frac{1}{2 i}  \int_{-\infty}^0 d\tau_1 \int_0^{\infty} d \tau_2  \left(  \left< \mathcal{O}_\alpha(\tau_1) \mathcal{O}_\beta(\tau_2)\right> -   \left< \mathcal{O}_\alpha(\tau_1) \mathcal{O}_\beta(\tau_2)\right>  \right) \nonumber \\  &=& \frac{1}{2 i}  \int_{-\infty}^0 d\tau_1 \int_0^{\infty} d \tau_2  \left< \left[ \mathcal{O}_\alpha(\tau_1), \mathcal{O}_\beta(\tau_2) \right]\right>. 
\end{eqnarray}

In summary
\begin{equation} \label{eq:tmcdc}
g_{\alpha \beta} = \int_{-\infty}^0 d\tau_1 \int_0^{\infty} d \tau_2 \left[ \frac{1}{2} \left< \left\lbrace \mathcal{O}_\alpha(\tau_1), \mathcal{O}_\beta(\tau_2) \right\rbrace_+\right> - \left< \mathcal{O}_\alpha(\tau_1) \right> \left< \mathcal{O}_\beta(\tau_2)\right> \right],
\end{equation}
\  and
\begin{equation} \label{eq:cbdc}
F_{\alpha \beta} =  \frac{1}{ i}  \int_{-\infty}^0 d\tau_1 \int_0^{\infty} d \tau_2  \left< \left[ \mathcal{O}_\alpha(\tau_1), \mathcal{O}_\beta(\tau_2) \right]\right>. 
\end{equation}
The above definitions are well known and can be used for any kind of quantum theory. 

\section{The Quantum Geometric Tensor for a Perturbed Theory}
Let us consider the case of a harmonic oscillator with a perturbation of $\lambda V(q)$, in this case, the Hamiltonian takes the form
\begin{equation}
H=\frac{1}{2}p^2 + \frac{\alpha}{2} q^2 + \lambda V(q),
\end{equation}
we will choose as parameters $\lambda^1 = \alpha$ and $\lambda^2=\lambda$. If we make the variation $\lambda^i \rightarrow \lambda^i + \delta \lambda^i$, we notice that 
\begin{equation}
H(\lambda^i+\delta \lambda^i) = H(\lambda^i) + \frac{\delta \alpha}{2}q^2 + \delta \lambda V(q),
\end{equation}
and therefore
\begin{equation}
\mathcal{O}_\alpha = -\frac{q^2}{2}, \ \ \ \  \ \ \mathcal{O}_\lambda =- V(q),
\end{equation}
then, using equation (\ref{eq:defQGT}), the components of the QGT for this theory are
\begin{equation} \label{eq:QGTaaN}
G_{\alpha \alpha} = \frac{1}{4} \int_{-\infty}^0 d\tau_1 \int_0^\infty d\tau_2 \left[ \left< q^2(\tau_1) q^2(\tau_2) \right> - \left< q^2(\tau_1) \right> \left<q^2(\tau_2) \right> \right],
\end{equation}
\begin{equation}\label{eq:QGTalN}
G_{\alpha \lambda} = \frac{1}{2} \int_{-\infty}^0 d\tau_1 \int_0^\infty d\tau_2 \left[ \left< q^2(\tau_1) V(q(\tau_2)) \right> - \left< q^2(\tau_1) \right> \left<V(q(\tau_2)) \right> \right],
\end{equation}
and
\begin{equation}\label{eq:QGTllN}
G_{\lambda \lambda } =  \int_{-\infty}^0 d\tau_1 \int_0^\infty d\tau_2 \left[ \left<V(q(\tau_1)) V(q(\tau_2)) \right> - \left< V(q(\tau_1)) \right> \left<V(q(\tau_2)) \right> \right].
\end{equation}
Since the theory with a  potential $V(q)$ is not solvable in general, we need to use perturbation theory to compute the QGT . With this in mind, equations (\ref{eq:QGTaaN})-(\ref{eq:QGTllN}) do not seem to give a real advantage to do calculations. However, as we will see in the next section, the path integral approach gives a robust process for computing the expectation values needed.

\section{Perturbative Approach to Green's functions}
Let us assume that we have a Lagrangian with the form
\begin{equation}
L(q,\dot{q})=\frac{1}{2}\left( \dot{q}^2-\alpha q^2 \right)-\lambda  V(q),
\end{equation}
then we make a Wick rotation and define the generating function $Z[J]$ as  
\begin{equation}
Z[J]=\int \mathcal{D}q(\tau)e^{-\int d \tau \left[ {\mathcal L}_E + J(\tau) q(\tau) + \lambda V(q) \right]},
\end{equation}
where ${\mathcal L}_E$ is the Euclidean Lagrangian of the free harmonic oscillator theory, i.e.
\begin{equation}
{\mathcal L}_E=\frac{1}{2}\left(\left( \frac{dq}{d\tau}\right)^2 +\alpha q^2 \right).
\end{equation}

We notice that
\begin{align}\label{41}
\frac{\delta^n Z[J]}{\delta J(\tau_1) ...\delta J(\tau_n)} = &(-1)^n\int \mathcal{D} q(\tau) q(\tau_1)...q(\tau_n) \nonumber \\
&\times \exp\left[-\int d\tau \left(  {\mathcal L}_E + J(\tau) q(\tau) + \lambda V(q) \right) \right],
\end{align}
then, we write the expectation values as
\begin{equation} \label{eq:expgreen}
\left< q(\tau_1)...q(\tau_n)\right>= (-1)^n \left. \left( \frac{1}{Z[J]} \frac{\delta^n Z[J]}{\delta J(\tau_1)...\delta J(\tau_n)}\right) \right|_{J=0}\equiv G^{int}_n(\tau_1,\dots,\tau_n).
\end{equation}
Then, using (\ref{41}), we rewrite  the Green's functions $ G^{int}_n(\tau_n...\tau_1)$ in terms of the action 	of the system $S=\int d\tau \left(   {\mathcal L}_E  + \lambda V(q) \right)$ as
\begin{equation} \label{eq:gfinact}
G_n^{int}(\tau_1,...,\tau_n)=\frac{\int \mathcal{D}q(\tau)q(\tau_1)...q(\tau_n)e^{-S[q(\tau)]}}{\int \mathcal{D}q(\tau)e^{-S[q(\tau)]}},
\end{equation}
the exponential of the action can be expressed as
\begin{equation}
e^{-S}=e^{-S_0-S_1}=e^{-S_0}\left[1+\sum_{m=1}^\infty \frac{(-1)^m}{m!}S_1^m \right],
\end{equation}
where $S_0= \int d\tau   {\mathcal L}_E$ and $S_1=\int d\tau  \lambda V(q) $. If we substitute in equation (\ref{eq:gfinact}) we obtain
\begin{equation} \label{eq:gfacc}
G_n^{int}(\tau_1,...,\tau_n)=\frac{\int \mathcal{D}q(\tau)q(\tau_1)...q(\tau_n)e^{-S_0[q(\tau)]}\left[1+\sum_{m=1}^\infty \frac{(-1)^m}{m!}S_1^m \right]}{\int \mathcal{D}q(\tau)e^{-S_0[q(\tau)]}\left[1+\sum_{m=1}^\infty \frac{(-1)^m}{m!}S_1^m \right]}.
\end{equation}
Assuming that the potential can be expanded in the form
\begin{equation}
V(q) = \sum_{n=0}^\infty  c_n q^n,
\end{equation}
where $c_n$ are certain constants. For simplicity, hereinafter we will suppose that the potential has the form
\begin{equation}
V(q)= \frac{q^k}{k!},
\end{equation}
Then,  we can rewrite equation (\ref{eq:gfacc}) in terms of the Green's functions of the simple harmonic oscillator
\begin{align} \label{eq:greenfunctcompser}
&G^{int}_n(\tau_1,\dots,\tau_n)=\nonumber \\
&\frac{G_n(\tau_1,\dots,\tau_n)+\sum_{m=1}^\infty \frac{(-\lambda/k! )^m}{m!}\int ds_1 ...ds_m G_{n+mk}(\tau_1,...,\tau_n,s_1^k,...,s_m^k)}{1+\sum_{m=1}^\infty \frac{(- \lambda/k! )^m}{m!}\int ds_1...ds_m G_{mk}(s_1^k,...,s_m^k)},
\end{align}
where we have defined 
\begin{equation}\label{Greenfunction1}
G^{int}_{n+m}(\tau_1^n,\tau_2^m) \equiv \left< q^n(\tau_1)q^m(\tau_2) \right>,  
\end{equation}
and similarly for functions including powers of $\tau$. 
In the next section we present some examples in which we take advantage of formulas (\ref{eq:QGTaaN})-(\ref{eq:QGTllN}) and (\ref{eq:greenfunctcompser}).
\section{Examples}
\subsection{Linear perturbation}
As as a first example we will consider the harmonic oscillator with a linear perturbation, i.e $V=q$. The Hamiltonian of this system is

\begin{equation}
H=\frac{1}{2}p^2+\frac{\alpha}{2}q^2+Jq,
\end{equation}
where we have set the coupling constant $\lambda=J$ to agree with standard quantum field theory notation. For its simplicity, this system can be solved exactly. Actually, the ground state for the system is
\begin{equation}
\Psi=\left( \frac{\sqrt{\alpha}}{\pi}\right)^{1/4} \exp\left(- \frac{\sqrt{\alpha}}{2} \left(q+\frac{J}{\alpha} \right)^2 \right),
\end{equation}
then the corresponding derivatives are
\begin{equation}
\frac{\partial \Psi}{\partial J}=-e^{-\frac{\sqrt{\alpha}}{2}\left(q +\frac{J}{\alpha} \right)^2}\frac{\left(J+q \alpha\right)}{\pi^{1/4}\alpha^{11/8}},
\end{equation}
and
\begin{equation}
\frac{\partial \Psi}{\partial \alpha}=\frac{e^{-\frac{\sqrt{\alpha}}{2}\left(q +\frac{J}{\alpha} \right)^2}\left(6J^2+4Jq\alpha+\alpha^{3/2}-2q^2\alpha^2 \right)}{8\pi^{1/4}\alpha^{19/8}}.
\end{equation}
We can now compute the connections and the components of the metric
\begin{equation}
\left< \partial_\alpha \Psi \right| \left. \Psi \right>=\left< \partial_J \Psi \right| \left. \Psi \right>=0,
\end{equation}
\begin{equation}
\left< \partial_J \Psi \right| \left. \partial_J \Psi \right> = \frac{1}{2 \alpha^{3/2}},
\end{equation}
\begin{equation}
\left< \partial_J \Psi \right| \left. \partial_\alpha \Psi \right>=-\frac{J}{2 \alpha^{5/2}},
\end{equation}
\begin{equation}
\left< \partial_\alpha \Psi \right| \left. \partial_\alpha \Psi \right>=\frac{1}{32 \alpha^2} + \frac{J^2}{2 \alpha^{7/2}},
\end{equation}
with this products we can easily find that the QGT components are
\begin{equation}
G_{\alpha \alpha}= \frac{1}{32 \alpha^2} + \frac{J^2}{2 \alpha^{7/2}},
\end{equation}
\begin{equation}
G_{J,\alpha}=-\frac{J}{2 \alpha^{5/2}},
\end{equation}
and
\begin{equation}
G_{J J}= \frac{1}{2 \alpha^{3/2}}.
\end{equation}
\subsubsection{Lagrangian approach}
Now we proceed to use our approach, first we notice from the Hamiltonian that
\begin{equation}
\mathcal{O}_\alpha = - \frac{q^2}{2}, \ \ \ \mathcal{O}_J =-q.
\end{equation}
Before we compute the Quantum Information Metric, we show how we calculate the expectation values for the Euclidean time in this specific case. The partition function for this problem is
\begin{eqnarray}
Z[J]   = \int \mathcal{D}q \exp\left[ - \int_{-\infty}^\infty d\tau \left( \frac{1}{2} \left( \frac{dq}{d\tau}\right)^2 +\frac{\alpha}{2}q^2+J q\right)\right], \label{eq:fun}
\end{eqnarray}
in the present problem, $J$ is a constant, but if we suppose $J=J(\tau)$ we can use the standard trick to compute the euclidean expectation values, i.e.
\begin{equation}
\left<q(\tau_1)...q(\tau_n) \right> =  \left( -1 \right)^n \left.\left( \frac{1}{Z[J]} \frac{\delta}{\delta J(\tau_n)}...\frac{\delta}{\delta J(\tau_1)} Z[J]\right)\right|_{J=const}.
\end{equation}
We must emphasize that in the usual case the derivatives are evaluated in $J=0$, which gives the expectation values for the free  harmonic oscillator, whereas in this case we are evaluating $J=constant$ to obtain the expectation values for the problem with a linear perturbation. The equation (\ref{eq:fun}) can be done and takes the form
\begin{equation} \label{eq:funcpartint}
Z[J]= Z[0] \exp\left[\frac{1}{2}\int d\tau_1 d\tau_2 J(\tau_1)D(\tau_1,\tau_2)J(\tau_2) \right],
\end{equation}
where
\begin{eqnarray}
D(\tau_1,\tau_2)=&\frac{1}{2 \sqrt{\alpha}}\left(\theta(\tau_1-\tau_2)\exp(- \sqrt{\alpha}(\tau_1 -\tau_2))+ \theta(\tau_2-\tau_1)\exp(\sqrt{\alpha}(\tau_1 -\tau_2))\right) \nonumber  \\ 
&=\frac{1}{2 \sqrt{\alpha}} e^{-\sqrt{\alpha}|\tau_1-\tau_2|},
\end{eqnarray}
is the solution of the equation \begin{equation}
\left( \partial_\tau^2 -\alpha\right)D(\tau,\tau')=-\delta(\tau-\tau'),
\end{equation}
with these results we can now compute the QGT.
\subsubsection{Component $G_{JJ}$}
The component $G_{J J}$ takes the form
\begin{eqnarray}
G_{J J}= \int_{-\infty}^0 d\tau_1 \int_0^\infty d\tau_2 \left(\left< q(\tau_1) q(\tau_2) \right> - \left<q(\tau_1) \right>\left< q(\tau_2)\right> \right),
\end{eqnarray}
For computing this element, we need the first and second derivatives of  (\ref{eq:funcpartint})
\begin{equation}
\frac{\delta Z}{\delta J(\tau)}=Z\int ds J(s) D(s,\tau),
\end{equation}
\begin{equation}
\frac{\delta^2 Z}{\delta J(\tau_2)\delta J(\tau_1)}=Z\left[ \int ds_1 ds_2 J(s_1)J(s_2)D(s_1,\tau_1)D(s_2,\tau_2)+D(\tau_1,\tau_2)\right],
\end{equation}
therefore
\begin{equation}
\left< q(\tau)\right> =- J \int ds  D(s, \tau),
\end{equation}
and
\begin{equation}
\left< q(\tau_1) q(\tau_2)\right> = J^2 \int ds_1 ds_2 D(s_2, \tau_2)D(s_1,\tau_1)+ D(\tau_1,\tau_2),
\end{equation}
then
\begin{equation}
\left< q(\tau_1) q(\tau_2)\right>-\left< q(\tau_1)\right>\left< q(\tau_2)\right>=D(\tau_1,\tau_2)= \frac{1}{2 \sqrt{\alpha}}  e^{- \sqrt{\alpha}(\tau_2-\tau_1)},
\end{equation}
 after integrating on $\tau_1$ and $\tau_2$, we find that
\begin{equation}
 G_{JJ}= \frac{1}{2 \alpha^{3/2}}.
 \end{equation} 
 
\subsubsection{Component $G_{\alpha J}$} 
 
 For the component $G_{\alpha J}$ we need
 \begin{eqnarray}
G_{\alpha J}=\frac{1}{2}\int_{-\infty}^0 d\tau_1 \int_0^\infty d\tau_2 \left( \left< q^2(\tau_1) q(\tau_2) \right> -\left< q^2(\tau_1) \right>\left< q(\tau_2) \right> \right).
 \end{eqnarray}
 In this case we need the third derivative
 \begin{eqnarray}
 & \frac{\delta^3 Z}{\delta J(\tau_3) \delta J(\tau_2) \delta J(\tau_1)}=Z\left[ \int ds_1 ds_2 ds_3 J(s_1)J(s_2)J(s_3)D(s_1,\tau_1)D(s_2,\tau_2)D(s_3,\tau_3) \right. \nonumber \\ &\left. +\int ds J(s)\left[ D(s,\tau_3)D(\tau_1,\tau_2)+D(s,\tau_2)D(\tau_1,\tau_3)+D(\tau_2,\tau_3)D(s,\tau_1)\right]  \right],
   \end{eqnarray}
   then
   \begin{eqnarray}
      \left< q^2(\tau_1) q(\tau_2)\right>=-Z\left[ J^3 \int ds_1 ds_2 ds_3 D(s_1,\tau_1)D(s_2,\tau_1)D(s_3,\tau_2)\right. \nonumber \\ \left. + J\int d s \left(D(s,\tau_2)D(0)+2 D(s,\tau_1)D(\tau_1,\tau_2) \right)\right],
   \end{eqnarray}
       where $D(0)$ stands for $D(\tau,\tau)$. Therefore
   \begin{eqnarray}
   \left< q^2(\tau_1) q(\tau_2)\right> - \left<q^2(\tau_1) \right> \left< q(\tau_2)\right>=-2J\int ds D(s, \tau_1)D(\tau_1,\tau_2) \\ =-\frac{J}{2\alpha}\int ds e^{-\sqrt{\alpha}|s-\tau_1|}e^{-\sqrt{\alpha}(\tau_2-\tau_1)},
      \end{eqnarray}
after integration and multiplying the corresponding $\frac{1}{2}$, we obtain
 \begin{equation}
 G_{\alpha J}= -\frac{J}{2 \alpha^{5/2}}.
 \end{equation}

 \subsubsection{Component $G_{\alpha \alpha}$}
 
 Finally, for the term $G_{\alpha \alpha}$ we need
 \begin{eqnarray}
  G_{\alpha \alpha}= \frac{1}{4}\int_{-\infty}^0 d \tau_1 \int_0^\infty d\tau_2 \left( \left< q^2(\tau_1) q^2(\tau_2) \right> - \left<q^2(\tau_1) \right> \left< q^2(\tau_2)\right> \right),
   \end{eqnarray}
   the corresponding functional derivative is
   \begin{eqnarray}
      &&\frac{\delta^4 Z}{\delta J(\tau_4) \delta J(\tau_3) \delta J(\tau_2) \delta J(\tau_1)}=\nonumber \\ &&Z\left[\int ds_1 ds_2 ds_3 ds_4 J(s_1) J(s_2)J(s_3) J(s_4) D(\tau_1,s_1)D(\tau_2,s_2)D(\tau_3,s_3) D(\tau_4,s_4) \right. \nonumber \\ &&+\int ds_1 ds_2 J(s_1)J(s_2)\left[ D(s_1,\tau_3)D(\tau_1,\tau_2)D(s_2,\tau_4)+D(s_1,\tau_2)D(\tau_1,\tau_3)D(s_2,\tau_4)\right. \nonumber \\ &&+ \left. D(\tau_2,\tau_3)D(s_1,\tau_1)D(s_2,\tau_4) + D(\tau_1,\tau_4)D(s_1,\tau_2)D(s_2,\tau_4)+D(s_1,\tau_1)D(\tau_4,\tau_2)D(s_2,\tau_3) \right. \nonumber \\ && \left. + D(s_1,\tau_1)D(s_2,\tau_2)D(\tau_4,\tau_3) \right] \nonumber \\ && \left. +D(\tau_4,\tau_3)D(\tau_1,\tau_2)+D(\tau_4,\tau_2)D(\tau_1,\tau_3)+D(\tau_2,\tau_3)D(\tau_4,\tau_1)\right],
      \end{eqnarray}
      from which we obtain
      \begin{eqnarray}
            \left< q^2(\tau_1) q^2(\tau_2) \right>=J^4\int ds_1 ds_2 ds_3 ds_4 D(s_1,\tau_1)D(s_2,\tau_1)D(s_3,\tau_2)D(s_4,\tau_2) \nonumber \\ +J^2 \int ds_1 ds_2 \left[ 2 D(0)D(s_1,\tau_1)D(s_2,\tau_2)+4D(\tau_1,\tau_2)D(s_1,\tau_1)D(s_2,\tau_2)\right] \nonumber \\ + D^2(0) +2 D^2(\tau_1,\tau_2).
      \end{eqnarray}
 Now we can write
      \begin{eqnarray}
            \left< q^2(\tau_1)q^2(\tau_2)\right> -\left<q^2(\tau_1) \right> \left< q^2(\tau_2)\right> =4J^2\int ds_1 ds_2 D(\tau_1,\tau_2)D(s_1,\tau_1)D(s_2,\tau_2) \nonumber \\ +2 D^2(\tau_1,\tau_2) \nonumber \\ =\frac{J^2}{2 \alpha^{3/2}}\int ds_1 ds_2 e^{-\sqrt{\alpha}(\tau_2-\tau_1)}e^{-\sqrt{\alpha}|s_1-\tau_1|}e^{-\sqrt{\alpha}|s_2-\tau_2|}+\frac{e^{-2\sqrt{\alpha}(\tau_2-\tau_1)}}{2 \alpha},
            \end{eqnarray}
   after the corresponding integrations and the factor $1/4$, we find that
\begin{equation}
G_{\alpha \alpha}=\frac{1}{32 \alpha^2} + \frac{J^2}{2 \alpha^{7/2}}.
\end{equation}  
In this way we have shown that the results obtained by the Lagrangian procedure correspond to the results obtained using the standard quantum computation.

\subsection{Anharmonic Oscillator}
%%%%%%%%%%%%%%%%%%%%%%%%%%%%%%%%%%%%%%%%%%%%
Now we set $V=\frac{q^4}{4!}$, and the problem becomes in the quartic anharmonic oscillator, whose Lagrangian is given by 

\begin{equation}
L = \frac{1}{2}\left( \dot{q}^2 - \alpha q^2 \right) - \lambda \dfrac{q^4}{4 !}.
\end{equation}
We must recall that the kernel of the Schr\"odinger operator has been computed in closed form in \cite{Slobodenyuk}, see also \cite{quartanosc}. Therefore, in principle, we could use this expression,   but taking into account that we need no only the kernel but also the Green's functions  it is more useful to compute the Quantum Geometric Tensor perturbatively. In this case, we will take $\lambda^1=\alpha$ and $\lambda^2=\lambda$.
If we make the substitution $\lambda^i \to \lambda^i
+\delta\lambda^i$ we find that
\begin{equation}
L(\lambda+\delta\lambda)=L(\lambda)-\frac{q^2}{2} \delta \alpha -\frac{q^4}{4!}\delta \lambda, 
\end{equation}
so
\begin{equation}
O_\alpha = - \frac{q^2}{2},
\end{equation}
and
\begin{equation}
O_\lambda = -\frac{q^4}{4!}.
\end{equation}
Since both operators are only polynomials in $q$, the quantum metric tensor will only involve terms like
\begin{equation}
\left< q^n(\tau_1) q^m(\tau_2) \right>,
\end{equation}
we can see from equation (\ref{eq:expgreen}) that such terms are proportional to the Green's functions of the system (\ref{Greenfunction1}).
In the next sections  we compute the elements of the quantum geometric tensor $G_{ij}$.
\subsubsection{Component $G_{\alpha \alpha}$}
The component $G_{\alpha \alpha}$ is, by definition
\begin{eqnarray}
G_{\alpha \alpha} =\frac{1}{4} \int_{-\infty}^0 d\tau_1 \int_0^\infty d\tau_2 \left( \left< q^2(\tau_1) q^2(\tau_2) \right> - \left<q^2(\tau_1) \right> \left< q^2(\tau_2) \right> \right) \nonumber \\ = \frac{1}{4} \int_{-\infty}^0 d\tau_1 \int_0^\infty d\tau_2 \left(G^{int}_4(\tau_1^2,\tau_2^2)-G^{int}_2(\tau_1^2)G^{int}_2(\tau_2^2) \right). \label{88.1}
\end{eqnarray}
Using Eq. (\ref{eq:greenfunctcompser}), we expand to order $\lambda$ the above expression and obtain
\begin{equation} \label{eq:g2l}
G_2^{int}(\tau^2)= G_2(\tau^2)  + \frac{\lambda}{4!} \int ds \left[G_2(\tau^2)G_4(s^4)-G_6(\tau^2,s^4) \right],
\end{equation}
and
\begin{equation} \label{eq:g4l}
G^{int}_4(\tau^2_1, \tau^2_2) = G_4(\tau_1^2,\tau_2^2) -\frac{\lambda}{4!}\int ds \left[ G_8(\tau_1^2,\tau_2^2,s^4) -G_4(\tau_1^2,\tau_2^2)G_4(s^4) \right],
\end{equation}
therefore, to order $\lambda$ the term between parenthesis in (\ref{88.1})  results
\begin{align}
&G_4^{int}(\tau_1^2,\tau_2^2)-G_2^{int}(\tau_1^2)G_2^{int}(\tau_2^2) = 2 G_2^2(\tau_1,\tau_2) \nonumber \\  &- \frac{\lambda}{4!}\int ds \left[24(2 G_2(\tau_1,s)G_2(\tau_2,s)G_2(0)G_2(\tau_1,\tau_2)+G_2^2(\tau_1,s)G_2^2(\tau_2,s)) \right],
\end{align}
here $G_2(0)=G_2(\tau,\tau)$.
\begin{figure} 
  \centering
   \includegraphics[width=0.7\textwidth]{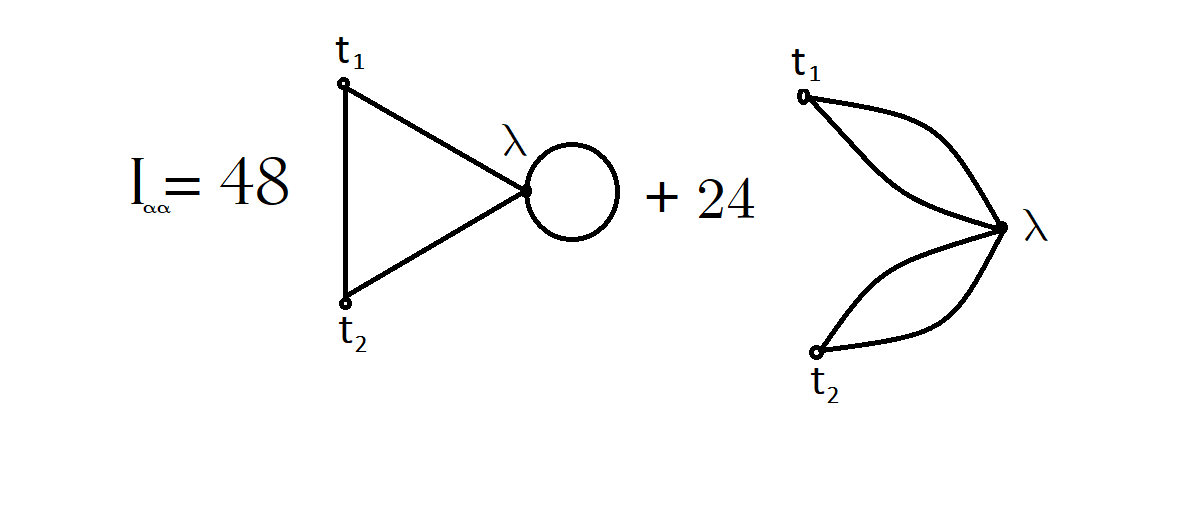}
  \caption{Integrand for computing the component $G_{\alpha \alpha}$ of the QGT.} \label{fig:aa}\end{figure}

If we call the integrand $I_{\alpha \alpha}$,
\begin{equation}
I_{\alpha \alpha}= 24(2 G_2(\tau_1,s)G_2(\tau_2,s)G_2(0)G_2(\tau_1,\tau_2)+G_2^2(\tau_1,s)G_2^2(\tau_2,s)),
\end{equation}
this term corresponds to the Feynman diagrams given in the Figure (\ref{fig:aa}). We observe, that in the case of Quantum Field Theory these diagrams will produce divergences, and the QGT must be regularized and renormalized.
In our case these diagrams are finite. Then using
\begin{equation} \label{eq:greenfunct}
G_2(\tau_1,\tau_2)=\frac{1}{2 \sqrt{\alpha}}\exp^{-\sqrt{\alpha}|\tau_2-\tau_1|},
\end{equation}
and after integration, we obtain
\begin{equation}
G_{\alpha \alpha}=\frac{1}{32 \alpha^2}- \frac{11\lambda }{512 \alpha^{7/2}} + O(\lambda^2),
\end{equation}
\subsubsection{Component $G_{\lambda \lambda}$}
\begin{figure} 
  \centering
   \includegraphics[width=1\textwidth]{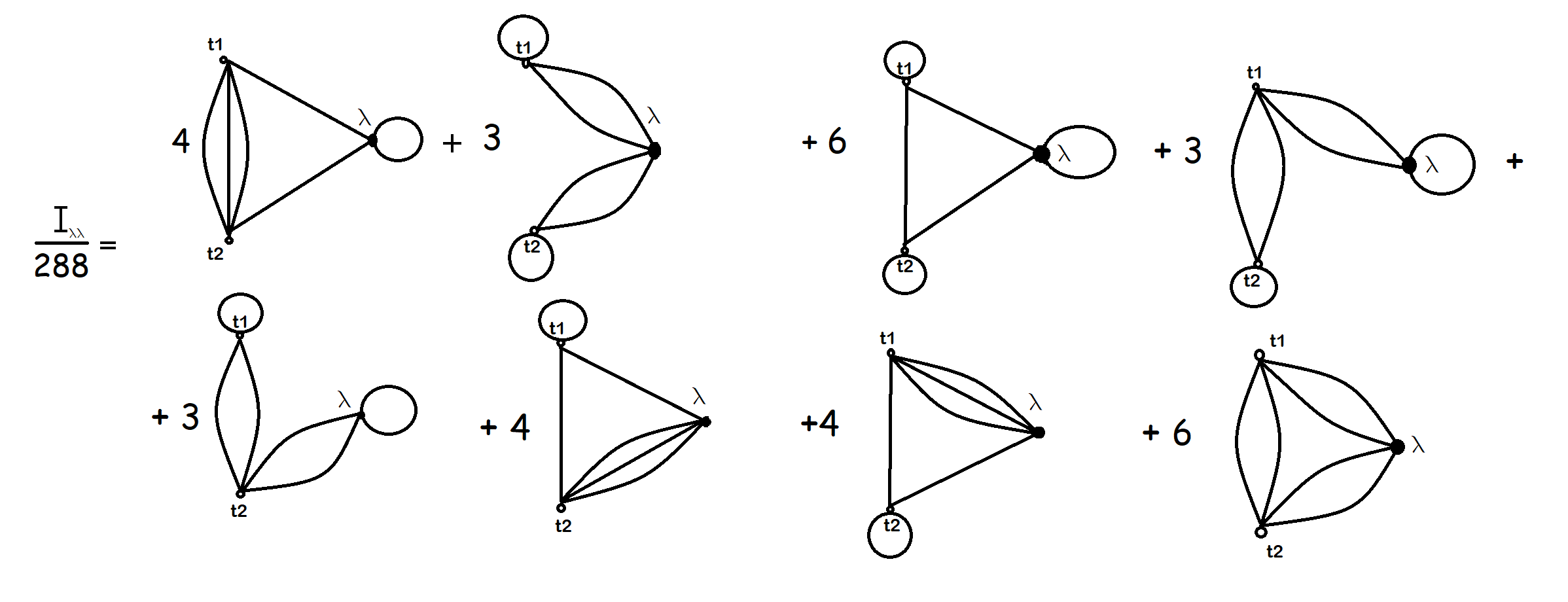}
  \caption{Integrand for computing the component $G_{\lambda \lambda}$ of the QGT.} \label{fig:ll}  \end{figure}

\begin{eqnarray}
G_{\lambda \lambda}= \frac{1}{(4!)^2} \int_{-\infty}^0 d\tau_1 \int_0^\infty d\tau_2 \left( \left< q^4(\tau_1)q^4(\tau_2)\right> -\left< q^4(\tau_1) \right> \left< q^4(\tau_2) \right>\right) \nonumber \\ = \frac{1}{(4!)^2} \int_{-\infty}^0 d\tau_1 \int_0^\infty d\tau_2 \left( G_8^{int}(\tau^4_1, \tau^4_2) - G^{int}_4(\tau^4_1) G^{int}_4(\tau^4_2)\right). \label{94.1}
\end{eqnarray}
To order $\lambda$
\begin{equation}
G^{int}_4(\tau^4) = G_4(\tau^4)-\frac{\lambda}{4!} \int ds \left[G_8(\tau^4, s^4)- G_4(\tau^4) G_4(s^2)\right],
\end{equation}
and
\begin{equation}
G_8^{int}(\tau_1^4,\tau_2^4)=G_8(\tau_1^4,\tau_2^4) - \frac{\lambda}{4!} \int ds G_{12}(\tau^4_1,\tau_2^4,s^4) + \frac{\lambda}{4!}\int ds G_4(s^4) G_8(\tau_1^4, \tau_2^4).
\end{equation}
%then
%\begin{eqnarray}
%G_8^{int}(t_1^4,t_2^4) -G_4^{int}(t_1^4)G_4^{int}(t_2^4)=G_8(t_1^4,t_2^4) -G_4(t_1^4)G_4(t_2^4) - \frac{\lambda}{4!} \int d\tau \left[G_{12}(t_1^4,t_2^4,\tau^4) \right. \nonumber \\ \left. -G_4(\tau^4)G_8(t_1^4, t_2^4)- %G_4(t_1^4)G_8(t_2^4,\tau^4)-G_4(t_2^4) G_8(t_1^4, \tau^4) + 2G_4(t_1^4)G_4(t_2^4)G_4(\tau^4) \right],
%\end{eqnarray}
Then integrand $I_{\lambda \lambda}$ in (\ref{94.1}) takes the form
\begin{eqnarray}
I_{\lambda \lambda}=288 \left[ 4 G_2(0)G_2^3(\tau_1,\tau_2)G_2(\tau_1,s)G_2(\tau_2,s)+3G_2^2(0)G_2^2(\tau_1,s) G_2^2(\tau_2,s) \right. \nonumber \\ \left.+ 6 G_2^3(0) G_2(\tau_1,\tau_2)G_2(\tau_1,s)G_2(\tau_2,s)+4G_2(0) G_2(\tau_1,\tau_2)G_2^3(\tau_1,s)G_2(\tau_2,s) \right. \nonumber \\ \left. +4 G_2(0) G_2(\tau_1,\tau_2)G_2(\tau_1,s)G_2^3(\tau_2,s)+3G_2^2(\tau_1,\tau_2)G_2^2(0)G_2^2(\tau_2,s)+\right. \nonumber \\ \left. +3G_2^2(\tau_1,\tau_2)G_2^2(\tau_1,s)G_2^2(0) + 6G_2^2(\tau_1,\tau_2)G_2^2(\tau_1,s)G_2(\tau_2,s)\right],
\end{eqnarray}
and we can see it in Figure (\ref{fig:ll}).
After integrating
\begin{equation}
G_{\lambda \lambda}=\frac{13}{6144 \alpha^3}-\frac{31 \lambda}{12288 \alpha^{9/2}} + O(\lambda^2).
\end{equation}
%%%%%%%%%%%%%%%%%%%%%%%%%
\subsubsection{Component $G_{\alpha \lambda}$}
\begin{figure} 
  \centering
   \includegraphics[width=0.7\textwidth]{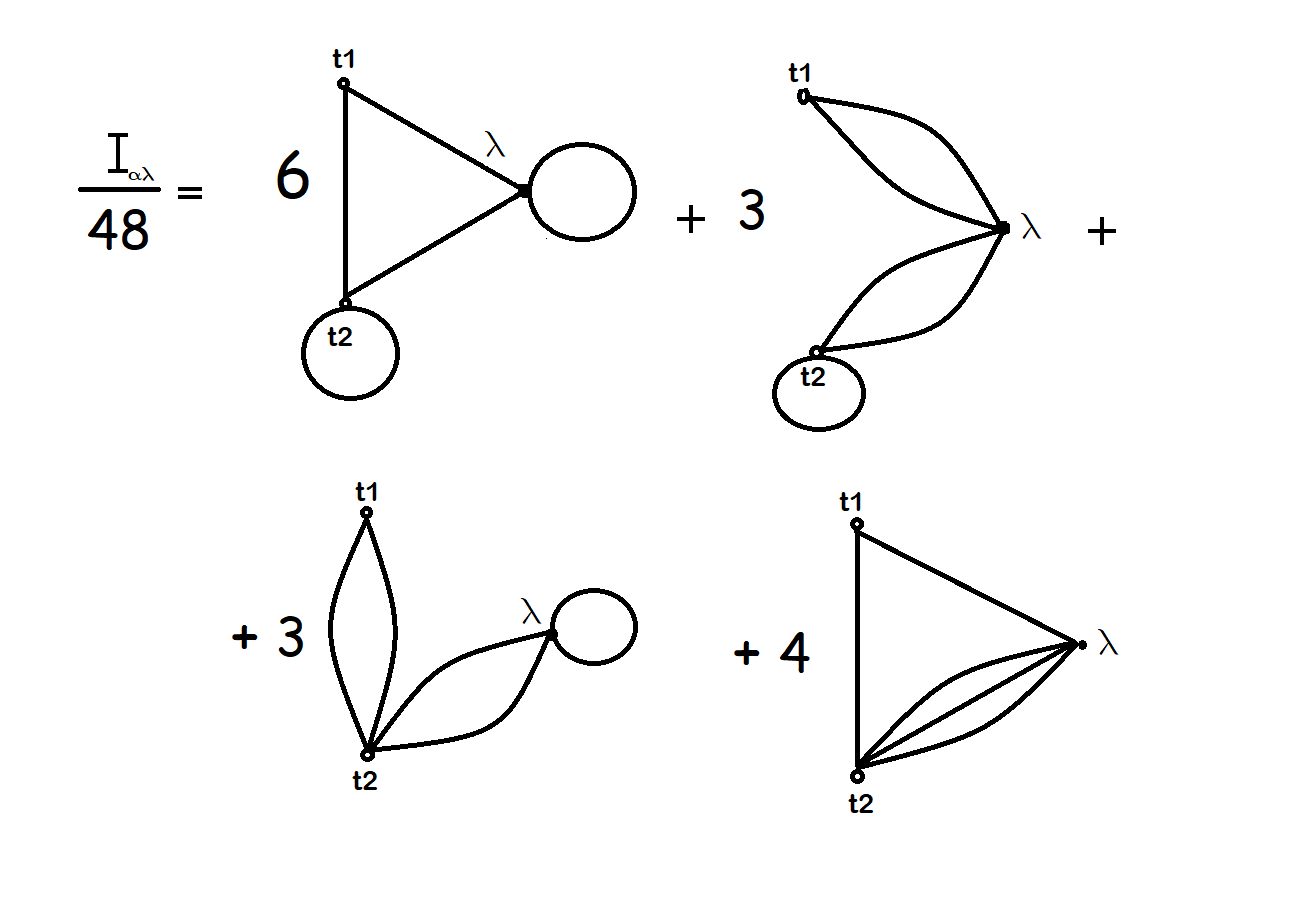}
  \caption{Integrand for computing the component $G_{\alpha \lambda}$ of the QGT.} \label{fig:al}\end{figure}

\begin{eqnarray}
G_{\alpha \lambda}= \frac{1}{2! 4!}\int_{-\infty}^0 d\tau_1 \int_0^\infty d\tau_2 \left[ \left< q^2(\tau_1)q^4(\tau_2) \right> -\left< q^2(\tau_1)\right> \left<q^4(\tau_2)\right> \right] \nonumber \\ =\frac{1}{2! 4!}\int_{-\infty}^0 d\tau_1 \int_0^\infty d\tau_2 \left[G^{int}_6(\tau_1^2, \tau_2^4)-G_2^{int}(\tau_1^2)G^{int}_4(\tau_2^4) \right].
\end{eqnarray}
For this element, we need the the functions (\ref{eq:g2l}) and (\ref{eq:g4l}). We can also expand $G_6^{int}(\tau_1^2,\tau_2^2)$
\begin{equation}
G_6^{int}(\tau_1^2, \tau_2^4)=G_6(\tau_1^2,\tau_2^4)- \frac{\lambda}{4!} \int ds \left[ G_{10}(\tau_1^2, \tau_2^4,s^4) - G_6(\tau_1^2,\tau_2^4)G_4(s^4) \right],
\end{equation}
and then
\begin{align}
G^{int}_6(\tau_1^2,\tau_2^4)-G_2^{int}(\tau_1^2)G_4^{int}(\tau_2^4) =G_6(\tau_1^2,\tau_2^4)-G_2(\tau_1^2)G_4(\tau_2^4)-\frac{\lambda}{4!} \int ds I_{\alpha \lambda},
\end{align}
where
\begin{eqnarray}
I_{\alpha \lambda}=48\left[ 6G_2(\tau_2,s)G_2^2(0)G_2(\tau_1,\tau_2)G_2(\tau_1,s)+3G_2(0)G_2^2(\tau_1,\tau_2)G_2^2(\tau_2,s) \right. \nonumber \\ \left.+3G_2(0) G_2^2(\tau_1,s) G_2^2(\tau_2,s)+4G_2(\tau_1,\tau_2)G_2(\tau_1,s) G_2^3(\tau_2,s)\right],
\end{eqnarray}
and we can see it in Fig. (\ref{fig:al}).
After integration
\begin{equation}
G_{\alpha \lambda}=\frac{1}{128 \alpha^{5/2}}- \frac{89 \lambda}{12288 \alpha^4} +O(\lambda^2).
\end{equation}
In this form we have computed all the components of the QIM for the quartic anharmonic oscillator using only Green's functions of the harmonic oscillator and this clearly shows the usefulness of our procedure.

An interesting conclusion is that given the QIM we can compute the determinant and we get to order $\lambda$,

\begin{equation}
\det(G_{ij}) = \frac{1}{196608 \alpha^5}- \frac{35 \lambda}{3145728 \alpha^{13/2}}+\mathcal{O}(\lambda^2).
\end{equation}
We note that the determinant becomes singular when
\begin{equation}
\lambda_c = \frac{16}{35}\alpha^{3/2} ,
\end{equation}
and this means that the QIM is degenerated for this value of the parameters and could indicate the validity limit of our approach.

\section{Conclusions}

In this work, we have shown that the Quantum Information Metric  (QIM)  could be computed by using the known Green's functions of the system. The procedure was shown, in first place by computing the QIM for the harmonic oscillator with a linear perturbation. In this case, the wave function, as well as the Green's function are well known. Therefore we were able to compute the results with the standard method and the one proposed here, and both ways produce the same results. In this form, if we can exactly compute the Green's functions  $G^{int}_n(\tau_1,\dots, \tau_n)$, we have the procedure to calculate the Berry curvature and the QIM in exact form.  Some additional examples of this procedure are shown in \cite{paper}. In the present work, we proceed in an alternative way using a perturbative method to compute the  Green's functions, this method was exemplified with the anharmonic oscillator with a perturbation of $q^4$. In this case, the quantum information metric was computed by calculating the Green's functions perturbatively to a given order in the coupling constant. This result shows that the quantum information metric can be computed perturbatively even when the wavefunction is not available at all. That result invites to calculations in field theory, where the wavefunction is usually not known, but there are several techniques for computing Green's functions perturbatively. We must mention that our procedure, in this case, is limited to small coupling and valid only when the metric is positive semidefinite and it is useful only to compute the  ground state QIM and Berry curvature. Furthermore, in our expressions, it is not possible to take the limit of the parameters to zero since this implies a divergent behavior of the metric.

\section*{Acknowledgments}

The authors acknowledge partial support from DGAPA-UNAM grants IN 103716, IN103919 and CONACyT project 237503. J. A. is supported by CONACyT scholarship 419420. We thank the referee for useful comments that helped us to improve our manuscript.

\end{document}